\documentclass[12pt,preprint]{aastex}

\usepackage{graphicx}
\usepackage{epstopdf}

\newcommand{\HI}{H$\,${\sc i}}

\newcommand{\sighi}{$\Sigma_{\rm HI}$}

\begin{document}

\title{A Search for correlations between turbulence and star formation in THINGS galaxies}

\author{Bruce G.\ Elmegreen\altaffilmark{1}, Zorayda Martinez\altaffilmark{2,3}, Deidre
A.\ Hunter\altaffilmark{2}}

\altaffiltext{1}{IBM T.\ J.\ Watson Research Center, 1101 Kitchawan Road, Yorktown Heights, New York USA}
\altaffiltext{2}{Lowell Observatory, 1400 West Mars Hill Road, Flagstaff, Arizona 86001, USA}
\altaffiltext{3}{Department of Astronomy and Planetary Science, Northern Arizona University, 527 S.\ Beaver Street, Flagstaff, Arizona 86011}

\begin{abstract}
The spatial range for feedback from star formation varies from molecular cloud
disruption on parsec scales to supershells and disk blowout on kiloparsec
scales. The relative amounts of energy and momentum given to these scales is
important for understanding the termination of star formation in any one region
and the origin of interstellar turbulence and disk stability in galaxies as a
whole. Here we measure for eleven THINGS galaxies the excess kinetic energy,
velocity dispersion and surface density of \HI\ gas associated with regions of
excess star formation, where the excess is determined from the difference
between the observed local value and the azimuthal average.  We find small
decreases in the excess kinetic energy and velocity dispersion in regions of
excess star formation rate density, suggesting that most of the feedback energy
does not go into local \HI\ motion. Most likely it disrupts molecular clouds and
dissipates rapidly at high gas density. Some could also be distributed over
larger regions, filling in spaces between the peaks of star formation and
contributing to other energy sources from self-gravity and spiral arm shocks.
\end{abstract}

\section{Introduction} \label{sec-intro}

Energy from massive stars in the form of ionization, radiation pressure, winds,
and supernovae drives gas expansion and turbulence in the neighborhoods of star
formation, disrupting the associated molecular clouds \citep[e.g.,][]{chevance20}
and powering lower density gas around them \citep[e.g.,][]{nath20}. An important
quantity is the fraction of turbulence this feedback powers.  Other potential
sources of turbulence have been known for a long time. They include disk
instabilities driven by gravity
\citep{devega96,bertin01,gammie01,huber01,wada02,vollmer02,elmegreen03,krumholz10},
vertical and in-plane instabilities driven by magnetism
\citep{parker65,asseo78,balbus91,sellwood99,kim03,piontek07}, thermal
instabilities \citep{hennebelle07,gazol10}, galactic shear
\citep{richard99,semelin00}, and energy coming from outside the disk, such as gas
accretion \citep{tenoriotagle81,elmegreenburkert10}, and galaxy interactions
\citep{elmegreen93,kaufman97,goldman00,burkhart10,ashley13,renaud14}. General
reviews on the origins of interstellar turbulence are in \cite{elmegreen04} and
\cite{maclow04}.

If a large fraction of interstellar turbulence is driven by young stellar
feedback, then star formation may regulate itself by inflating the disk and
lowering the density when the star formation rate is high
\citep{goldreich65,franco84,ostriker10}. If feedback is unable to power a high
fraction of the disk, then large-scale turbulence and marginal disk stability may
need another driver, such as self-gravity \citep{li05}, which gets more active
inversely with the turbulent speed via the Toomre $Q$ parameter
\citep{kim02,kim07}.

A combination of these processes with different fractions in different places is
likely. Some observations suggest that gravity or magnetic instabilities dominate
turbulence generation at a low star formation rate density (SFRD),
$\leq10^{-9}\;M_\odot$ pc$^{-2}$ yr$^{-1}$, while feedback dominates at high SFRD
\citep{agertz09,tamburro09}. Models by \cite{kim13} suggest that feedback controls
turbulence and the SFRD in the outer disk where the SFRD is low. The observations
for local galaxies can be ambiguous. An observed correlation between gas velocity
dispersion and SFRD could be the result of correlations between each quantity and
galactocentric radius \citep[e.g.,][]{stilp13a} with no mutual correlation between
them. Also, the lack of a correlation between turbulence and SFRD could be the
result of sampling the SFRD at the wrong time \citep{stilp13b}.  Feedback could
also dominate turbulence on small scales while self-gravity dominates it on large
scales \citep{joung09}.

Numerical simulations of turbulence in an interstellar patch show that supernovae
and other types of feedback can give the observed velocity dispersion and scale
height \citep{norman96,avila01,deavillez05,dib06}. This does not necessarily mean
that the large-scale velocity dispersion varies with SFRD. \cite{joung06} found
that most of the feedback energy is deposited within a few hundred pc of the
energy source. \cite{joung09} also show that even with only feedback to excite the
gas, the mass-weighted velocity dispersion (which is similar to the kinetic energy
density discussed below) and the simulated \HI\ linewidths do not change much with
SFRD when higher SFRDs correspond to higher gas surface densities according to the
Kennicutt-Schmidt relation. Moreover, turbulence in an interstellar patch cannot
capture larger-scale processes like gravitational instabilities \citep{balbus99}
or spiral shocks unless they are included specifically, and then these processes
may dominate turbulence driving, as found by \cite{kim06,kim10}. For example,
spiral shocks in M51 have enormous peculiar speeds, 50 km s$^{-1}$ in some regions
\citep{shetty07}, suggesting that these shocks are a good source of turbulent
motions.

If disk gravity maintains $Q\sim1$ through spiral instabilities and gas collapse
into clouds, then the turbulent speed is partly defined by that condition, i.e.,
it depends on the effective mass surface density and epicyclic frequency
\citep[e.g.,][]{goldbaum16}.  Simulations by \cite{bournaud10} modeled this case
to fit the infrared dust power spectrum of the Large Magellanic Cloud and pointed
out that feedback was necessary primarily to prevent the accumulation of gas in
very dense clumps, which was a runaway process with only gravity present.
\cite{hopkins11} also found that the disruption of dense gas was the most
important role of feedback and that without it, the SFR would be higher by a
factor of 100. \cite{hennebelle14} also find that supernovae clustered in regions
of star formation lower the star formation rate by a factor of 30 compared to
random supernovae, emphasizing again the importance of dense gas disruption in
feedback control. \cite{combes12} simulated the power spectrum in M33 and also
noted that feedback primarily influenced the high-frequency regime and was
necessary to correctly get that part of the power spectrum and the associated
inflection point (from two-dimensional turbulence to three-dimensional
turbulence). \cite{walker14} reproduced the power spectra of galaxies in the
THINGS survey with high feedback models, as weak feedback gave too shallow a power
spectrum, i.e., too much small scale power.

More recent simulations of THINGS galaxy power spectra by \cite{grisdale17} show
that feedback can influence a wide range of scales, up to 1-2 kpc, which is more
than the disk thickness. As in the previous work, too little feedback increased
the high frequency structure and flattened the power spectrum there. They also
point out that large-scale gravity is important too because without it, the power
spectrum is too shallow on large scales, i.e., there is too little large-scale
structure.

Evidently, both disk gravity and its activity prior to or independent of star
formation, plus feedback after star formation, are essential to reproduce observed
gas structures and motions throughout galaxy disks. Both may be required for the
regulation of star formation also, even though either one alone can give the right
star formation rate and turbulence speed with reasonable parameters. This blend of
distinct processes makes it difficult to find the boundary between them, whether
measured as the scale separating large-scale gravity and small-scale star
formation effects, or as the relative contribution of each to turbulence
generation and self-regulation.

In this paper we examine the excess \HI\ turbulence in localized regions of star
formation by removing the average radial profiles of each. Turbulence is measured
by the kinetic energy density, KED, and by the second moment of the \HI\ spectra. We
also measure excess \HI\ column density as a function of excess SFRD. The goal is to
estimate the fraction of star formation feedback energy that goes into \HI\ turbulence
locally. If this fraction is low, then this energy either spreads out from each
star formation site quickly so there is little excess energy density there, or the
energy is dissipated almost entirely in phases and scales of the interstellar
medium that are not observed with \HI, such as molecular clouds.

\section{Data}

\subsection{Galaxy Sample}\label{subsec:2.1}

The 11 galaxies used here, 10 spirals and one dIrr, were drawn from THINGS
\citep{walter08}, a large \HI\ survey of nearby galaxies using the Very Large
Array (VLA\footnote[1]{The VLA is a facility of the National Radio Astronomy
Observatory. The National Radio Astronomy Observatory is a facility of the
National Science Foundation operated under cooperative agreement by Associated
Universities, Inc.}). The THINGS \HI\ emission maps are available for download,
and we used the robust-weighted integrated \HI\ (moment 0) and velocity dispersion
(moment 2) maps. The pixel scale is $1.5\arcsec$ except for NGC 2403, which has a
pixel scale of $1.0\arcsec$.

To quantify the star formation rate density, we used far-ultraviolet (FUV) images
taken with the NASA Galaxy Evolution Explorer \citep[{\it GALEX},][]{galex}. These
images, obtained from the {\it GALEX} archives, were geometrically transformed and
smoothed to the \HI\ map orientation, pixel scale, and resolution so that the
images could be directly compared. The native resolution of GALEX FUV imagery is
$\sim4\arcsec$.

\begin{deluxetable}{ccccccccc}
\tabletypesize{\scriptsize}
\tablecaption{The Galaxy Sample  \label{table:1}}
\tablewidth{0pt}
\tablehead{
\colhead{}  &  \colhead{} & \colhead{D\tablenotemark{a}} & \multicolumn{2}{c}{\HI\ RO Beam\tablenotemark{b}} & \colhead{Inclination\tablenotemark{c}} & \colhead{PA\tablenotemark{d}} & \multicolumn{2}{c}{Center\tablenotemark{e}} \\
\colhead{Galaxy} & \colhead{Type} & \colhead{(Mpc)} & \colhead{Major (\arcsec)} & \colhead{Minor (\arcsec)} & \colhead{(\arcdeg)} & \colhead{(\arcdeg)} & \colhead{RA ($^{h}:^{m}:^{s}$)} & \colhead{Dec (\arcdeg:\arcmin:\arcsec)} \\
 }
 \startdata
     DDO 154 & dIrr & 3.7 & 7.94 & 6.27 & 65.2 & 46.0 & 12:54:06.25 & +27:09:02.0\\
     NGC 925 & Sd & 9.2 & 4.85  & 4.65 & 66.0 & 286.6 & 02:27:16.5 & +33:34:43.5 \\
     NGC 2403 & Scd & 3.2 & 6.01  & 5.17 & 62.9 & 123.7 & 07:36:51.1 & +65:36:02.9 \\
     NGC 2841 & Sb & 14.1 & 6.06 &  5.79 & 73.7 & 152.6 & 09:22:02.6 & +50:58:35.4 \\
     NGC 2976 & Sc & 3.6 & 5.25  & 4.88 & 64.5 & 334.5 & 09:47:15.3 & +67:55:00.0 \\
     NGC 3198 & Sc & 13.8 & 7.64  & 5.62 & 71.5 & 215.0 & 10:19:55.0 & +45:32:58.9 \\
     NGC 4736 & Sab & 4.7 & 5.96  & 5.55 & 41.4 & 296.1 & 12:50:53.0 & +41:07:13.2 \\
     NGC 5055 & Sbc & 10.1 & 5.78  & 5.26 & 59.0 & 101.8 & 13:15:49.2 & +42:01:45.3 \\
     NGC 6946 & Scd & 5.9 & 4.93  & 4.51 & 32.6 & 242.7 & 20:34:52.2 & +60:09:14.4 \\
     NGC 7331 & Sb & 14.7 & 4.94  & 4.60 & 75.8 & 167.7 & 22:37:04.1 & +34:24:56.5 \\
     NGC 7793 & Sd & 3.9 & 10.37 & 5.39 & 49.6 & 290.1 & 23:57:49.7 & -32:35:27.9 \\
\enddata
\tablenotetext{\textrm{a}}{Distance to the galaxy from \cite{walter08}.}
\tablenotetext{\textrm{b}}{The major and minor beam sizes as presented in \cite{walter08}.}
\tablenotetext{\textrm{c}}{The inclination of the galaxies in degrees \citep{deblok08}.}
\tablenotetext{\textrm{d}}{Position angle of the galaxy in degrees from \cite{deblok08}.}
\tablenotetext{\textrm{e}}{Center of the galaxy in RA and Dec as given in \cite{trach08}.}
\end{deluxetable}

The sample galaxies and basic properties are given in Table \ref{table:1}.

\subsection{Creating Images}\label{subsec:2.2}

For the {\it GALEX} FUV images, foreground and background objects were removed and
replaced with an average of the noise in a two pixel wide annulus around the
object using the {\scshape imedit} tool in the Image Reduction and Analysis
Facility \citep[IRAF,][]{iraf}. We used {\scshape imsurfit} to construct a
two-dimensional fit to the sky and subtract it from the cleaned FUV image. Pixels
outside of the galaxy determined by eye were blanked using the task {\scshape
blank} in the Astronomical Image Processing System (AIPS) developed by NRAO in
order to prevent noise from affecting the pixel-pixel comparisons.

The moment 0 (MOM0) map became the \HI\ mass surface density ($\Sigma_{\rm HI}$),
the moment 2 (MOM2) map became the velocity dispersion ($V_{\rm disp}$), and the
FUV image became the star formation rate per area, called here the star formation
rate density (SFRD). The kinetic energy density (KED) was constructed from $0.5
\times \Sigma_{\rm HI} \times V_{\rm disp}^2$. The units for the four quantities
are 10$^{43}$ ergs pc$^{-2}$ for KED, km s$^{-1}$ for $V_{\rm disp}$, M$_{\odot}$
pc$^{-2}$ for $\Sigma_{\rm HI}$, and 10$^{-10}$ M$_{\odot}$ pc$^{-2}$ yr$^{-1}$
for SFRD. The KED was corrected for the presence of He and heavy elements in the
galaxy using 1.34 $\times\;\Sigma_{\rm HI}$. The conversion factor used for
transforming the MOM0 map units of Jy beam$^{-1}$ m s$^{-1}$ to atoms cm$^{-2}$ is
given in Table \ref{table:2}. To convert FUV flux to a SFRD, we used $SFRD =
{FUV}/{10^{6.466}}$, derived from \citet{kennicutt98} for the spirals and for the
dwarf DDO 154 $SFRD = {FUV}/{10^{6.508}}$, as given by \citet{hunter10}.

The moment 2 value of the \HI\ line profile, rather than the FWHM, is used to
probe HI turbulence because MOM2 contains more information about the line-wings,
making it a more sensitive measure of excess motion.  MOM2 is also a better
measure of kinetic energy \citep{tamburro09}. We are looking for any indication
that star formation energizes the local atomic gas, and this could include high
speed motions in shells or other disturbances that appear in the line-wings.

High speed shear, compression and expansion from spiral density waves could also
contribute to MOM2 values inside the $\sim6^{\prime\prime}$ \HI\ beam. This
angular size corresponds to several hundred parsecs for the more distant galaxies.
These contributions to MOM2 will increase the measured velocity dispersions mostly
in the arms where there are spiral density wave shocks, but they should not
contribute much between the arms or in quiescent regions where the gas flows more
smoothly. As a result, the average value of MOM2 used for background subtraction
could be a little less than the background value in the arms, and thus the
difference between the MOM2 values in star-forming regions, which are mostly in
the arms, and the average background could be somewhat higher than it would be
without spiral arm streaming motions. For this reason, the excess $V_{\rm disp}^2$
and KED values associated with regions of excess star formation should be
considered upper limits to the effects of feedback. Because these excess $V_{\rm
disp}^2$ and KED are already low (see below), the feedback energy going into \HI\
should be even lower than what our results suggest.

The physical contributions to $V_{\rm disp}^2$ measured from MOM2 could also vary
from region to region or between galaxies. Some regions with locally high excess
$V_{\rm disp}^2$ could contain expanding \HI\ shells with significant feedback
energy into the local \HI. Other regions with low or negative excess $V_{\rm
disp}^2$ could have a different star formation age or another destination for the
stellar energy and momentum. Here we consider the average trends of excess KED and
$V_{\rm disp}^2$ versus excess SFRD as a measure of the global effects of feedback
on the local atomic medium.

\begin{deluxetable}{ccccc}
\tabletypesize{\scriptsize} \tablecaption{Constants    \label{table:2}}
\tablewidth{0pt} \tablehead{ \colhead{}  & \colhead{Mass Conversion
Factor\tablenotemark{a}} & \colhead{$\Sigma_{\rm HI}$
Calibration\tablenotemark{b}}
& \colhead{KED Calibration\tablenotemark{c}} & \colhead{$\Delta R$\tablenotemark{d}} \\
\colhead{Galaxy} & \colhead{($10^{19}$ atom cm$^{-2}$ (Jy beam$^{-1}$ m s$^{-1}$)$^{-1}$ )}
& \colhead{(M$_{\odot}$ pc$^{-2}$ (Jy beam$^{-1}$ m s$^{-1}$)$^{-1}$)}
& \colhead{($10^{43}$ ergs pc$^{-2}$)} & \colhead{(\arcsec)} \\
}
\startdata
     DDO 154 & 2.220 & 0.1780 & $2.371 \times 10^{-7}$ & 6 \\
     NGC 925 & 4.900 & 0.3926 & $5.232 \times 10^{-7}$ & 50 \\
     NGC 2403 & 3.556 & 0.2850 & $3.798 \times 10^{-7}$ & 60 \\
     NGC 2841 & 3.149 & 0.2524 & $3.363 \times 10^{-7}$ & 60 \\
     NGC 2976 & 4.313 & 0.3456 & $4.606 \times 10^{-7}$ & 50 \\
     NGC 3198 & 2.574 & 0.2062 & $2.748 \times 10^{-7}$ & 60 \\
     NGC 4736 & 3.341 & 0.2677 & $3.567 \times 10^{-7}$ & 50 \\
     NGC 5055 & 3.635 & 0.2912 & $3.881 \times 10^{-7}$ & 50 \\
     NGC 6946 & 4.970 & 0.3982 & $5.307 \times 10^{-7}$ & 50 \\
     NGC 7331 & 4.863 & 0.2294 & $3.057 \times 10^{-7}$ & 50 \\
     NGC 7793 & 1.977 & 0.1584 & $2.111 \times 10^{-7}$ & 70 \\
\enddata
\tablenotetext{\textrm{a}}{Conversion constant for converting the Moment 0 map
into units of atom cm$^{-2}$.} \tablenotetext{\textrm{b}}{Conversion factor that
puts the Moment 0 map into units of M$_{\odot}$ pc$^{-2}$.}
\tablenotetext{\textrm{c}}{A factor to convert the Moment 0 $\times$ (Moment
2)$^{2}$ maps into units of ergs pc$^{-2}$.} \tablenotetext{\textrm{d}}{Width of
the annuli used in determining the radial profiles as described in Section
\ref{subsec:2.3}.}
\end{deluxetable}

\clearpage
\subsection{Radial Profiles}\label{subsec:2.3}

We are interested in the relation between \HI\ turbulence and local star
formation.  Because there could be several sources of \HI\ turbulence, we consider
only the local excess MOM2 above its average radial value and examine this excess
as a function of the excess SFRD above the SFRD average radial value. If local
star formation drives local turbulence, then there should be a positive
correlation between these excess values. The \HI\ quantities come from maps that
have full coverage in each disk, even between and beyond the star-forming regions,
but the SFRD is too low to measure in some places as star formation is generally
patchy. We consider the \HI\ excesses only in regions where the SFRD can be
measured. Thus, the average radial values for \HI\ come from everywhere and are
representative of all the \HI\ as a function of radius, but the local excesses
above these averages come from the star-forming regions.

\begin{figure}[ht]
\includegraphics[width=12cm]{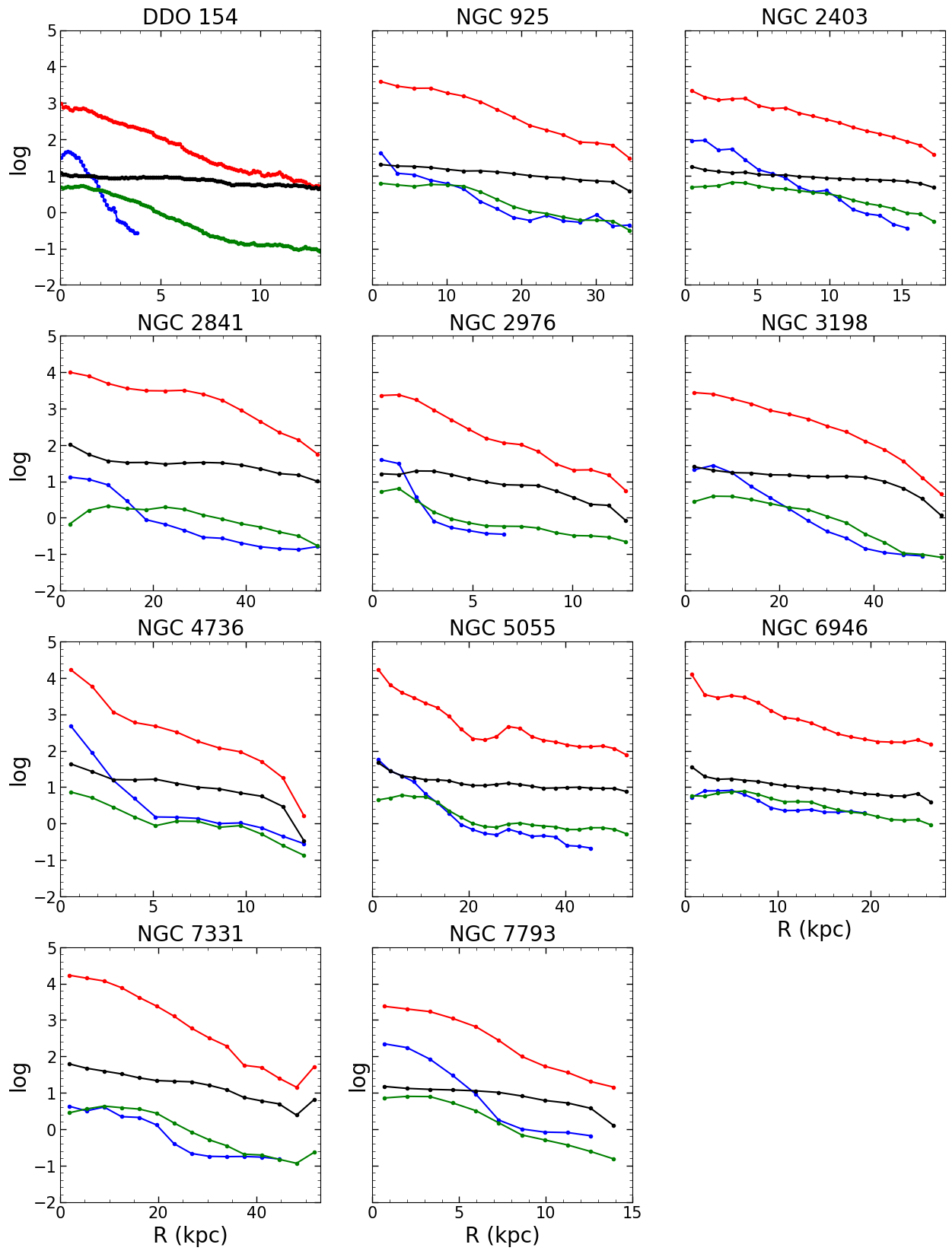}
\caption{Azimuthally-averaged radial profiles created as described in Section \ref{subsec:2.3}.
Red curves are log KED in units of
$10^{43}$ ergs pc$^{-2}$, black curves are log $V_{\rm disp}$ in units of km s$^{-1}$,
green curves are log \sighi\ in units of $M_{\odot}$ pc$^{-2}$, and blue curves are
log SFR/area in units of $10^{-10}\;M_{\odot}$ pc$^{-2}$ yr$^{-1}$. The plotted SFRD
is limited by the extent of the FUV. 
\label{allradaves_v2}}
\centering
\end{figure}

Determining the radial profiles of the sample involves superposing elliptical
annuli onto the images in order to cover the majority of the signal in both the
SFRD image and the images derived from the \HI\ maps. (Note that the \HI\ maps
extend significantly further than the FUV emission in all galaxies). The center of
each galaxy and the position angle of the major axis are given in Table
\ref{table:1} . The inclination of the galaxy was used to derive the
minor-to-major axis ratio $b/a$ of the ellipses, assuming that the intrinsic $b/a$
due to the thickness of the disk of spirals is 0.2 while that of dwarfs is 0.3
\citep{hodge66}. Approximately ten times the \HI\ beam-size was used as the width
of the annuli, $\Delta R$, which is given in Table \ref{table:2}. This gives
annuli wide enough to contain a statistically significant area, but not so large
that the annulus extends over a region where the exponential disk drops off
significantly. This judgement was made by eye. The exception to this was DDO 154
where we used an annulus width of $6\arcsec$ to match the radial profiles in
\citet{hunter12}. Quantities were also corrected to face-on by multiplying the
flux per pixel by the cosine of the inclination of the galaxy.  The resulting
average radial profiles are shown in Figure \ref{allradaves_v2}. There are no
obvious features corresponding to spiral arms in the optical disk. To determine
the excesses by subtracting the appropriate average from each pixel value, we
found the distance of each pixel from the center of the galaxy in the plane of the
disk and assigned this distance to the correct annulus.

\begin{figure}[ht]
\includegraphics[width=16cm]{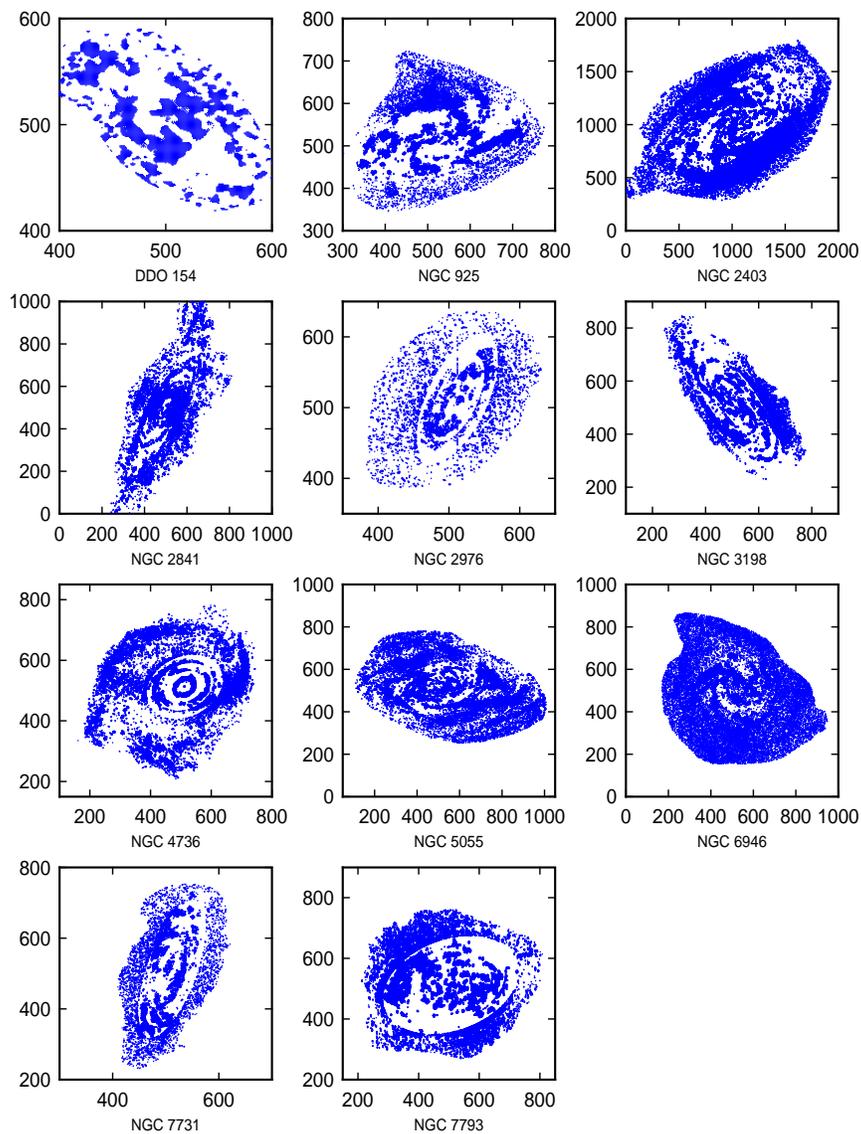}
\caption{Distributions of pixels with positive values of SFRD,
used for the determination of excess KED, $V_{\rm disp}$ and $\Sigma_{\rm HI}$ as
functions of excess SFRD. Units on the axes are pixels, which are $1.5\arcsec$ across for all
galaxies but NGC 2403, where they are $1.0\arcsec$. 
\label{fig:map}}
\centering
\end{figure}

\begin{figure}[ht]
\plotone{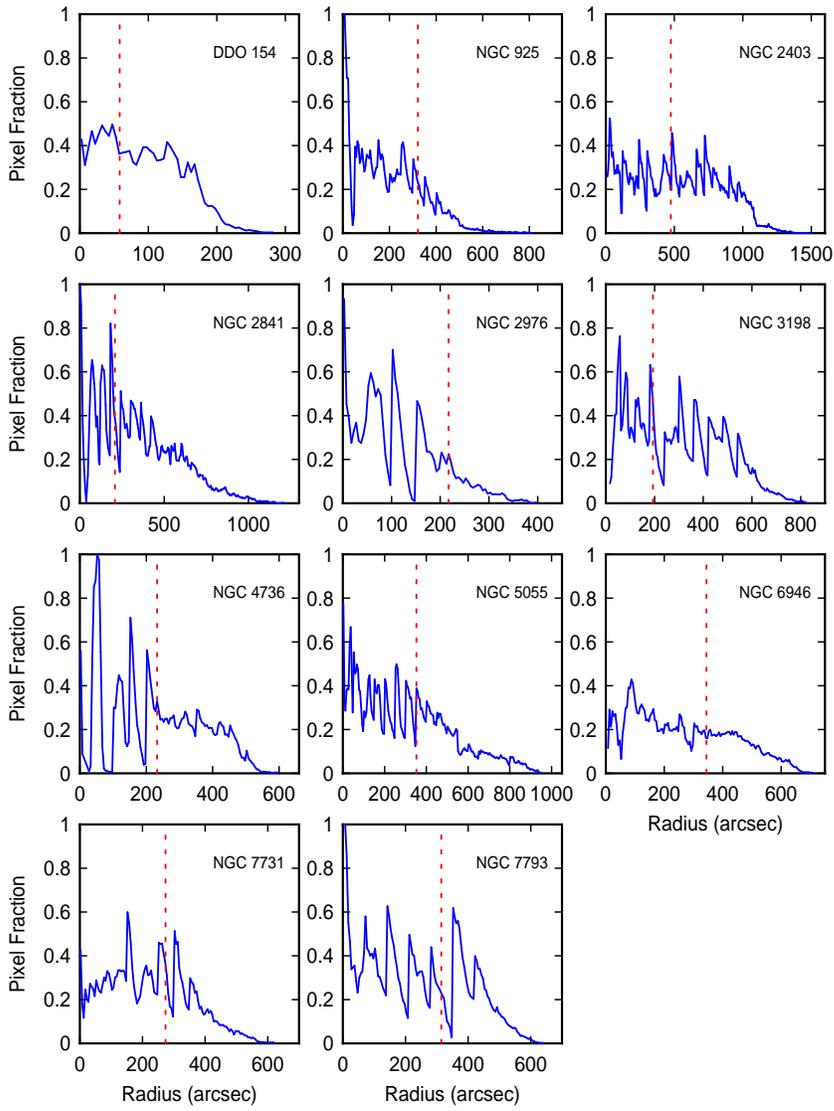}
\vskip -.9truein
\caption{The fraction of pixels as a function of radius that are used for the
calculation of the excess values. The vertical red dotted line is the isophotal
radius $R_{25}$.
\label{fig:pixelfraction_vs_radius}}
\centering
\end{figure}

Each annulus used to create the radial profile was checked for the number of SFRD
pixels. Towards the outer galaxy there may be many values of KED, $\Sigma_{\rm
HI}$, or $V_{\rm disp}$, but a statistically less significant number of values for
SFRD. In order to maintain a lower uncertainty in the averages, annuli that
contained less than 100 pixels in the SFRD were not plotted. Maps showing all of
the pixels with measured SFRD are shown in figure \ref{fig:map}; there are traces
of spiral structure because the SFRD is highest there. The fraction of pixels that
are included for the excess SRFD values is in figure
\ref{fig:pixelfraction_vs_radius}. This fraction ranges between 0.2 and 0.4 inside
$R_{\rm 25}$ and tapers off beyond 1.5 to 2 $R_{\rm 25}$.

\subsection{Pixel-pixel Plots}\label{subsec:2.4}

Each pair of values of an excess quantity in a pixel, such as the excess KED and
the corresponding excess SFRD, was plotted as a point in the two-dimensional plane
of these quantities. The density of these points on the plane shows the
probability distribution function for the correlation between values. We refer to
these plots as pixel-pixel plots. In constructing these plots, we first eliminated
pixels from the total KED, $\Sigma_{\rm HI}$, $V_{\rm disp}$, and SFRD maps with
values less than or equal to zero as they represent blanked pixels. Then, for each
of the non-zero pixels, the annular average value at that position determined from
all the pixels in the annulus, including the negative pixels, was subtracted from
the positive pixel value to give the excess value. We fit the resulting
pixel-pixel plots with a color density scale that shows the locations of highest
pixel densities. NGC 5055 is shown for illustration in Figure
\ref{pixel_plots_KED_vdisp_NHI}. The rest of the galaxies are shown in Appendix A.

Radial profiles of the average excess values are in figures
\ref{fig:ked_vs_radius} to \ref{fig:sfr_vs_radius}. The optical radius $R_{25}$ is
indicated by a vertical dotted red line.  The velocity dispersion excess generally
increases with radius, and the $\Sigma_{\rm HI}$ excess decreases a little with
radius, offsetting each other to make the KED excess about constant. The excess
SFRD decreases strongly with radius, like the average exponential disk itself,
suggesting that star-forming regions are selected to be at a fixed multiplicative
threshold above the average FUV disk. For this reason, pixels with large excess
SFRD in the pixel-pixel plots tend to be in the inner disks. The large excesses in
KED and $\Sigma_{\rm HI}$ at 260 arcsec in DDO 154 and 600 arcsec in NGC 5055 are
well beyond the optical radii.

There is sometimes a correspondence between the excursions in these figures. There
is a large bump at 100 arcsec in NGC 2976 for both excess $\Sigma_{\rm HI}$ and
KED, which probably corresponds to large \HI\ clouds in the southwest and
northeast, as shown in figure 25 of \cite{walter08}.  This bump is not present for
excess $V_{\rm disp}$ nor SFRD. This difference implies that the excess KED is
from the surface density part of this quantity, not the velocity dispersion part.
NGC 6946 also has a bump of excess KED and $\Sigma_{\rm HI}$ in the radial range
between 170 and 360 arcsec, and again there is no elevated excess $V_{\rm disp}$
there. This region corresponds to the end of the optical spiral arms where there
are giant \HI\ complexes \citep[fig. 65 in][]{walter08}. Neither of these features
in NGC 2976 nor NGC 6946 show up prominently in the average radial profiles in
figure \ref{allradaves_v2}; they are excesses relative to this average profile. In
about half of the galaxies, the excess SFRD flattens beyond $R_{\rm 25}$, but in
these cases, excess $V_{\rm disp}$ shows no indication of a different trend there.

Overall, the radial profiles of the excesses indicate that the correlations or
lack of correlations between turbulence generation and local star formation
discussed in this paper are for regions beyond the optical disk, which are still
relatively bright in FUV and \HI. Possible correlations for the main optical disk
will show up at the highest excess SFRD, exceeding around $-10$ in the log with
units of $M_\odot$ pc$^{-2}$ yr$^{-1}$, as indicated by the SFRD excesses to the
left of the vertical dotted lines in Figure \ref{fig:sfr_vs_radius}.

\section{Analysis}

We are interested in how much gas kinetic energy and turbulence each region of
star formation generates in its neighborhood on the scale of resolution of the
THINGS survey. As mentioned in the previous section, we removed large-scale
variations by subtracting the average radial profile of a quantity from the
individual pixel values of that quantity, referring to the result as the excess.
Figure \ref{pixel_plots_KED_vdisp_NHI} showed sample plots of excess \HI\ KED,
\HI\ $V_{\rm disp}$ and $\Sigma_{\rm HI}$ versus excess SFRD for one galaxy.

There is a lot of scatter in each plot, increasing with higher excess SFRD, but
the most common KED excesses and velocity dispersions are small for a wide range
of excess SFRD. To quantify these results, we determined four trend lines for each
plot. One, representing the most probable correlation, is the excess gas value for
each SFRD excess measured at the peak density of pixel points. Another is the rms
average of the excess gas value for each excess SFRD measured for all pixel points
above the most probable trend line, and a third is the rms average for all pixel
points below the most probable trend line. The fourth trend line is the difference
between rms above and below the most probable, added to the most probable. This
fourth trend is the upward bias from the most probable correlation, and is taken
to be indicative of a statistical upper limit of the quantity for each SFRD.

For example, in the case of a KED pixel plot, we made a histogram of the number of
points as a function of KED inside each narrow range, $\pm0.5$, of log excess
SFRD, where excess SFRD is measured in units of $M_\odot$ pc$^{-2}$ yr$^{-1}$ as
in the figures. The excess KED at the peak of the histogram was then determined.
Figure \ref{KED_vdisp_NHI_peaks} shows these most probable trend lines as solid
curves. The rms averages above and below the most probable trend lines were
determined in the usual way, from the square root of the difference between the
average square of the values and the square of the average of the values. These
rms averages are the long-dashed curves. The bias trend is shown by a short-dashed
curve.

Figure \ref{ked_vdisp_nhi_vs_sfra_6panel_with_log} shows the most probable curves
in the top panels and the biased or statistical upper limit curves in the bottom
panels for each galaxy. The excess KED bias values are mostly positive, so we plot
them in log scale in the top left panel with each galaxy labeled. The excess KED
values themselves hover around 0 and are plotted on a linear scale in the lower
left. The other panels are linear scale for the most probable and bias values.

The most probable excess KED and $V_{\rm disp}$ values have essentially no
dependence on excess SFRD, and are even a little negative for all SFRD, which
means that the KED and \HI\ $V_{\rm disp}$ decrease a little in each region of
star formation compared to the azimuthal average values. Upper limits to these
values, as shown by the bias curves in the bottom row of figure
\ref{ked_vdisp_nhi_vs_sfra_6panel_with_log}, have a clear upward trend for KED and
$\Sigma_{\rm HI}$ and a downward trend for $V_{\rm disp}$. Because KED is half the
product of $\Sigma_{\rm HI}$ and the square of $V_{\rm disp}$, the upward trend in
KED is entirely from the upward trend in $\Sigma_{\rm HI}$, not from increased
turbulent speeds. That is, localized star formation pushes around extra gas in its
vicinity, but at lower than average speed. This result applies for all galaxies
and SFRDs in the THINGS survey; it is independent of distance, and therefore not
likely to result from resolution limits, and independent of size.

From the bottom row of figure \ref{ked_vdisp_nhi_vs_sfra_6panel_with_log}, the
average slope of the log of the excess KED upper limits versus the log of the
excess SFRD is $0.17\pm0.11$. The average slope of the excess $V_{\rm disp}$ upper
limits versus the log of the excess SFRD is $-0.46\pm0.87$. The average slope of
the log of the excess $\Sigma_{\rm HI}$ upper limits (for positive values) versus
the log of the excess SFRD is $0.27\pm0.18$. The inverse of this latter quantity
is analogous to the Kennicutt-Schmidt relation for \HI\ gas, but here it is for
the excess quantities only. We derive
\begin{equation}d\log(\Sigma_{\rm SFR}-<\Sigma_{\rm SFR}>)/
d\log(\Sigma_{\rm HI}-<\Sigma_{\rm HI}>)=2.0\pm2.0,\end{equation}
including only positive values of $\Sigma_{\rm HI}$ excess from the upper limit curves,
which means omitting all of NGC 6946
which has only negative values. This slope is approximately the same, although with a
large uncertainty, as the slope
of the KS relation for total \HI\ found for THINGS galaxies by \cite{roy15},
also using FUV for the SFR; their slope was $1.65\pm0.04$ on a scale of 1 kpc.

The slope of the excess SFRD versus the excess $\Sigma_{\rm HI}$ could have
some bearing on the process of star formation. If we think of the Kennicutt-Schmidt relation
as the zero-order correlation in a disk galaxy, involving the bulk gas and the
SFRD averaged over many local regions, then an excess correlation as in
Figure \ref{ked_vdisp_nhi_vs_sfra_6panel_with_log}
could be different, showing instead
a first-order trend, which might be something relevant to the local rate inside the
local regions. Presumably the zero-order correlation involves the average rate at which
bulk gas collects together to make star-forming clouds, in which case a first-order correlation
might involve the average rate at which stars form inside the clouds once the clouds have formed.
We discussed elsewhere how the zero-order correlation should have a slope of around
1.5 if the gas thickness varies with radius in a galaxy more slowly than the disk surface
density, which is usually the case, making the gas thickness approximately constant
in the main disk \citep{elmegreen18,wilson19}. The correlation inside self-gravitating clouds
could have a different slope, such as $\sim2$, if the cloud thickness depends on its surface
density \citep{elmegreen18}. Perhaps the correlation between excess $\Sigma_{\rm HI}$
and excess SFRD is hinting at this distinction from the zero-order Kennicutt-Schmidt relation.

\section{Fraction of Star Formation Supernova Energy going into Turbulence}

Figure \ref{all_ked_over_KED_from_sfra_vs_sfra} shows the ratio of the excess KED upper limits
to the KED expected from 100\% of the supernova energy put out by the excess SFRD. This
maximum supernova energy is from \cite{bacchini20},
\begin{equation}
KED_{\rm SN} = \Sigma_{\rm SFR}f_{\rm cc}E_{\rm SN}(2H/v_{\rm turb})
\label{eq:bacchini}
\end{equation}
where $f_{\rm cc}=1.3\times10^{-2}\;M_\odot^{-1}$ is the number of core-collapse
supernovae per solar mass of stars, $E_{\rm SN}=10^{51}$ erg is the supernova
energy, $H$ is the disk thickness and $v_{\rm turb}$ is the turbulent velocity
dispersion (not the excess turbulent dispersion).
The ratio of the observed KED excess to the maximum KED from the excess SFRD is
the local-excess analogue of the efficiency $\eta$ in \cite{bacchini20}. The ratios are plotted versus excess
SFRD assuming fiducial values of $H=100$ pc and $v_{\rm turb}=10$ km s$^{-1}$.

Figure \ref{all_ked_over_KED_from_sfra_vs_sfra} shows a clear trend of $\eta$ decreasing with
increasing excess SFRD. The value $-2$ on the ordinate ($\eta=1$\%) is approximately what
\cite{bacchini20} got for the average required feedback efficiency, leading them to conclude
that there is enough
supernova energy to power interstellar turbulence. Here that value appears again
where the {\it excess} SFRD is about the same as the {\it average} SFRD in main galaxy disks, i.e.,
$10^{-9}\;M_\odot$ pc$^{-2}$ yr$^{-1}$, which is in agreement with \cite{bacchini20},
but we see systematically higher values
for lower excess SFRD.  The highest values of
$\eta$, greater than unity ($\log\eta>0$) at the lowest excess SFRD, indicate that the excess KED
upper limit is more than the excess star formation
can generate even at 100\% efficiency. This high value
is similar to what others get in the outer parts of disks where the SFRD is low
\cite[e.g.,][]{tamburro09} and it suggests there is an additional source of turbulence.
The essential origin of the trend in Figure \ref{all_ked_over_KED_from_sfra_vs_sfra} is that
excess KED is more constant than excess SFRD, so the ratio that appears in $\eta$ varies
as the inverse of the excess SFRD. As in Figure
\ref{ked_vdisp_nhi_vs_sfra_6panel_with_log},
the results in Figure \ref{all_ked_over_KED_from_sfra_vs_sfra} suggest that
star formation does not significantly influence the KED of local \HI.

The lack of a direct correlation between local SFRD and local KED or $V_{\rm disp}$
does not mean there is no feedback, but only that most of the feedback energy does not
significantly move local atomic gas. Most of it likely goes into the
dense molecular gas where it pumps in gravitational potential energy by
pushing apart cloud pieces and where it radiates efficiently. The corresponding expansion
of the associated \HI\ would then have a relatively low velocity because it is a minor
component of the mass. Some of the feedback could also get channeled
to remote regions via low-density cavities.
For feedback to regulate the SFRD by adjusting the gas scale height, a fraction such as
$\sim1.5$\% \citep{bacchini20} of all the feedback, not just the local excess,
would have to be distributed widely without any significant trace of
local or direct agitation by young stars.
Regulation of the star formation rate
by the break-up of star-forming clouds seems more plausible than adjusting the
galactic scale height, given the current results.

\section{Conclusions}

There is no correlation between excess \HI\ velocity dispersion or excess kinetic energy density
and the excess star formation rate per unit area in eleven THINGS galaxies, where excess
is defined to be the measured local value minus the azimuthal average value at that
position. This result implies that star formation observed in the FUV does not
generate significant turbulence in the nearby atomic gas.  Either the kinetic energy
and momentum generated in the environments of young stars spreads so rapidly over very large
regions that it does not show up locally, or this energy and momentum is deposited
in a phase of gas that does not show up in the \HI\ survey. Most likely,
a significant fraction of the feedback energy
from young stars goes into adjacent molecular clouds. Some could also go
into cool atomic clouds, such as the debris from shredded molecular clouds,
which also would not show up well in the second moment map used here for $V_{\rm disp}$
because of the narrower linewidths and slower motions of the cool clouds.

There is a slight cooling trend with increasing excess SFRD, in the sense that the
statistical upper limits to the excess velocity dispersions decrease systematically
with increasing SFRD for all galaxies. This trend could correspond to a decay of local
\HI\ turbulence before star formation begins in a region, when the gas is condensing
into molecular clouds because of self-gravity.

\acknowledgements

ZM appreciates funding from the National Science Foundation (grants 1852478 and 1950901)
to Northern Arizona University
for the 2021 Research Experiences for Undergraduates program. We are grateful to the
referee for useful comments.

Facilities: \facility{VLA} \facility{GALEX}

\newpage

\begin{figure}
\includegraphics[width=16.0cm]{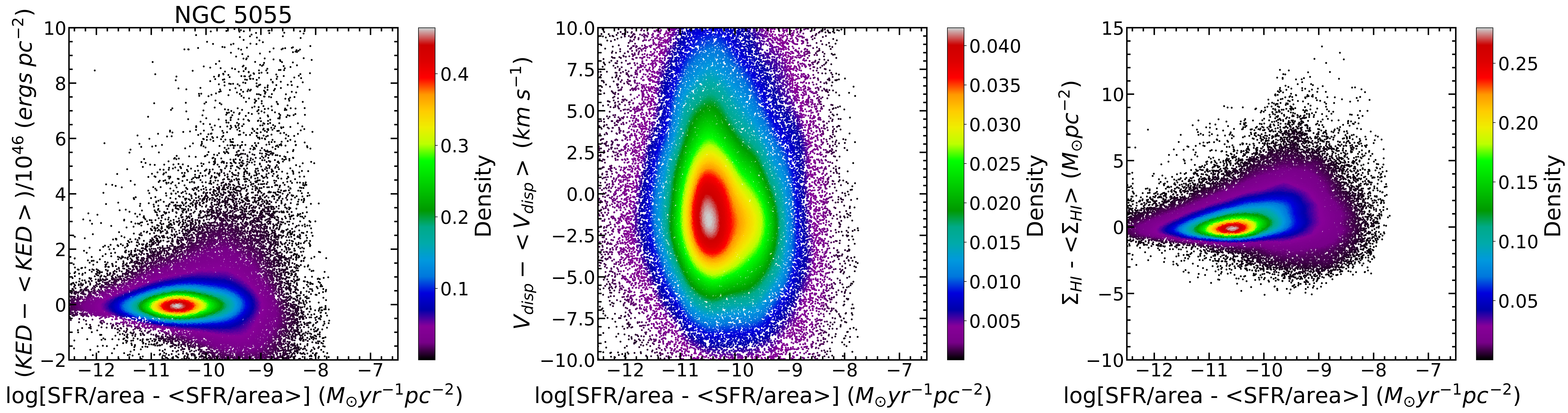}
\caption{Pixel-pixel plots showing the excess KED in units of $10^{46}$
erg pc$^{-2}$, $V_{\rm disp}$
in km s$^{-1}$, and $\Sigma_{\rm HI}$ in M$_{\odot}$ pc$^{-2}$
against the log of the excess SFRD in M$_{\odot}$
pc$^{-2}$ yr$^{-1}$.  The excess is defined to be the
difference between the local values and the values from the average radial profiles.
The color scale represents the density of points.
NGC 5055 is shown here for illustration; the other galaxies are shown in Appendix A.
\label{pixel_plots_KED_vdisp_NHI}}
\centering
\end{figure}

\begin{figure}[ht]
\plotone{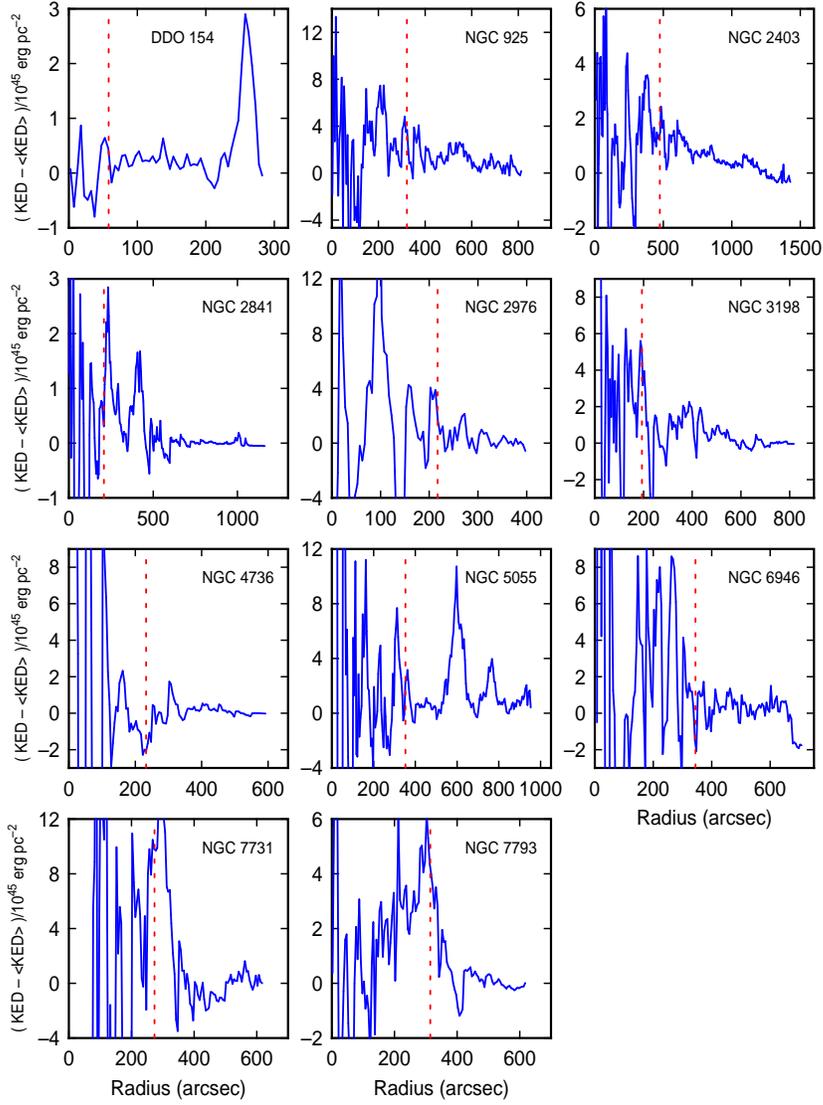}
\vskip -.9truein
\caption{The excess Kinetic Energy Density in \HI,
in units of $10^{45}$ erg pc$^{-2}$ is shown as a function of
galactocentric radius in arcsec. The vertical dotted line is the optical radius,
$R_{\rm 25}$.
\label{fig:ked_vs_radius}}
\centering
\end{figure}

\begin{figure}[ht]
\plotone{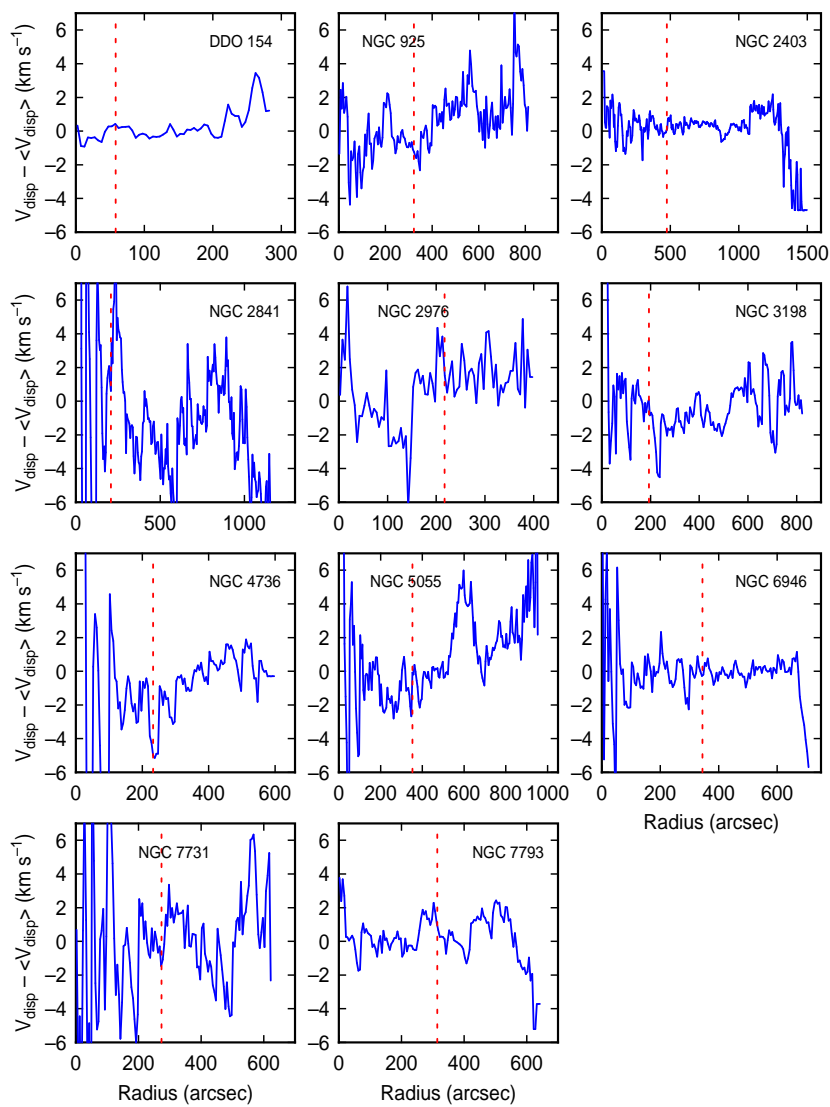}
\vskip -.9truein
\caption{The excess MOM2 velocity dispersion for \HI,
in units of km s$^{-1}$ is shown as a function of
galactocentric radius in arcsec. The vertical dotted line is the optical radius,
$R_{\rm 25}$.
\label{fig:v_vs_radius}}
\centering
\end{figure}

\begin{figure}[ht]
\plotone{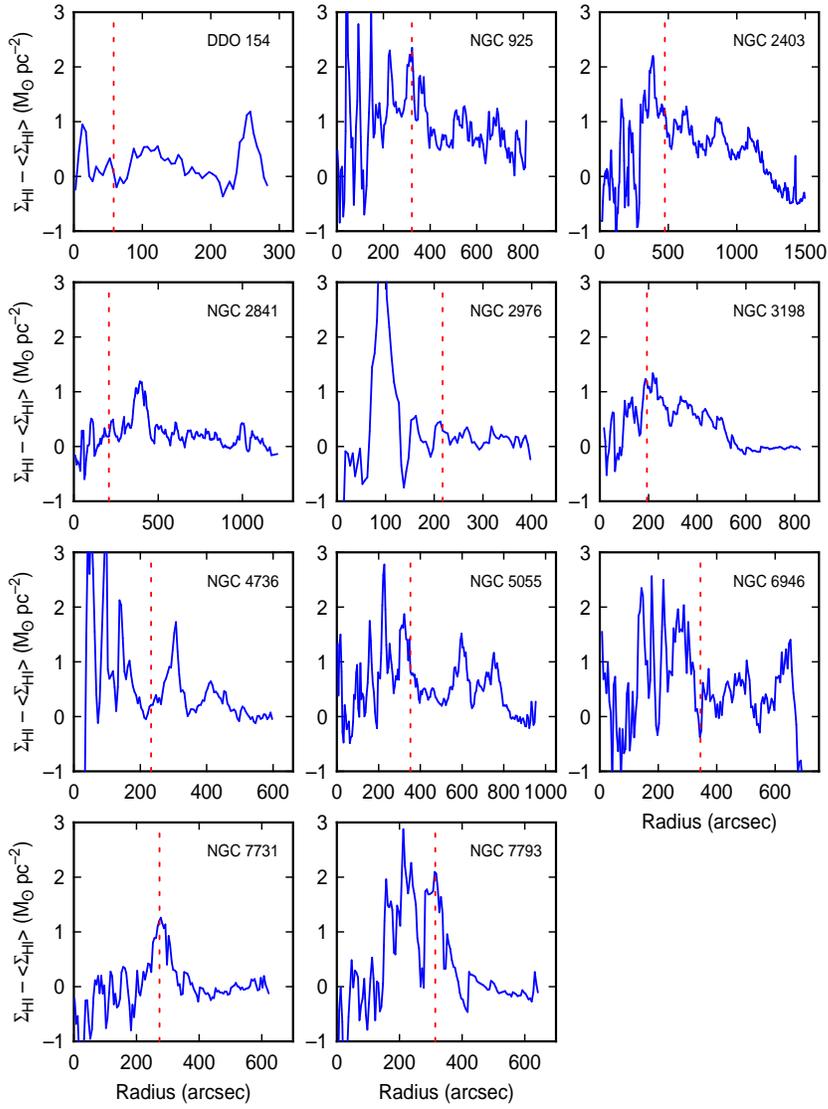}
\vskip -.9truein
\caption{The excess surface density of \HI\ in units of $M_\odot$ pc$^{-2}$
is shown as a function of
galactocentric radius in arcsec. The vertical dotted line is the optical radius,
$R_{\rm 25}$.
\label{fig:nhi_vs_radius}}
\centering
\end{figure}

\begin{figure}[ht]
\plotone{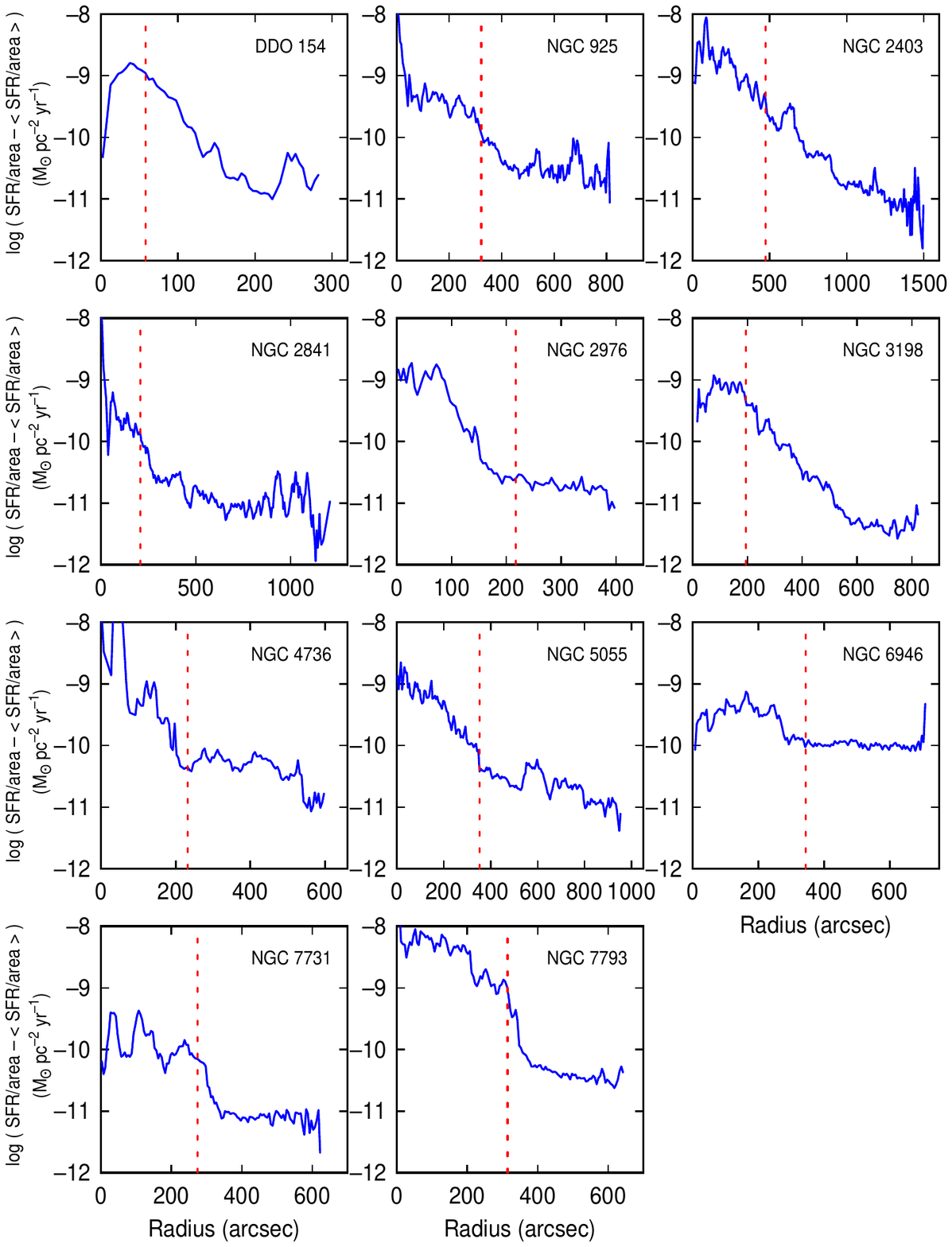}
\vskip -.9truein
\caption{The excess star formation rate density, SFR/Area, in units of M$_{\odot}$
pc$^{-2}$ yr$^{-1}$ is shown as a function of
galactocentric radius in arcsec. The vertical dotted line is the optical radius,
$R_{\rm 25}$.
\label{fig:sfr_vs_radius}}
\centering
\end{figure}

\newpage
\begin{figure}
\epsscale{1.}
\vskip -1.4truein
\plotone{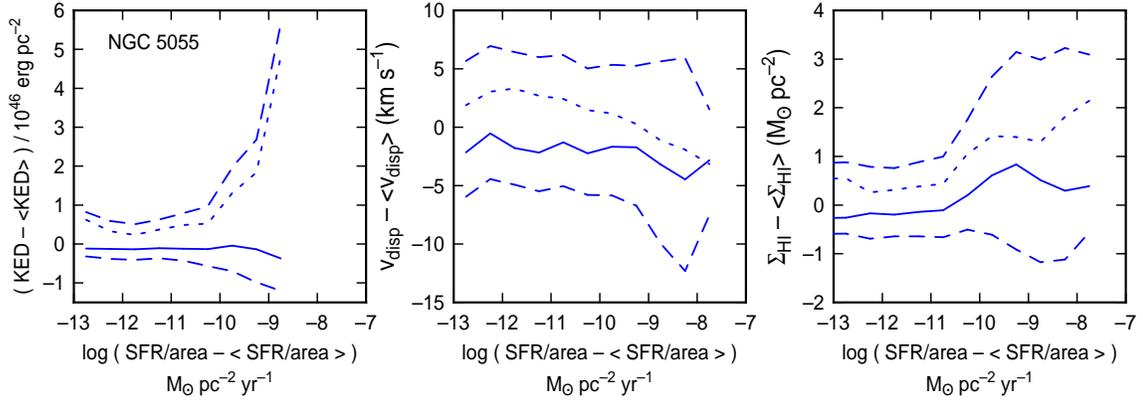}
\vskip -.7truein
\caption{Sample limits for the pixel distributions in Figure
\ref{pixel_plots_KED_vdisp_NHI}. The solid lines
represent the values of the plotted quantities at the
peak densities in the pixel plots, the dashed lines represent the
rms deviations of the plotted quantities from their values at the peak densities,
and the dotted line is the difference between the positive and negative rms values
added to the values at the peak. This dotted line is called the
upward bias in the main text, and taken to be the statistical upper limit.
The units for the quantities are the same as in Figure \ref{pixel_plots_KED_vdisp_NHI}.
\label{KED_vdisp_NHI_peaks}}
\end{figure}

\newpage
\begin{figure}[t!]
\epsscale{1.}
\vskip -3.5truein
\plotone{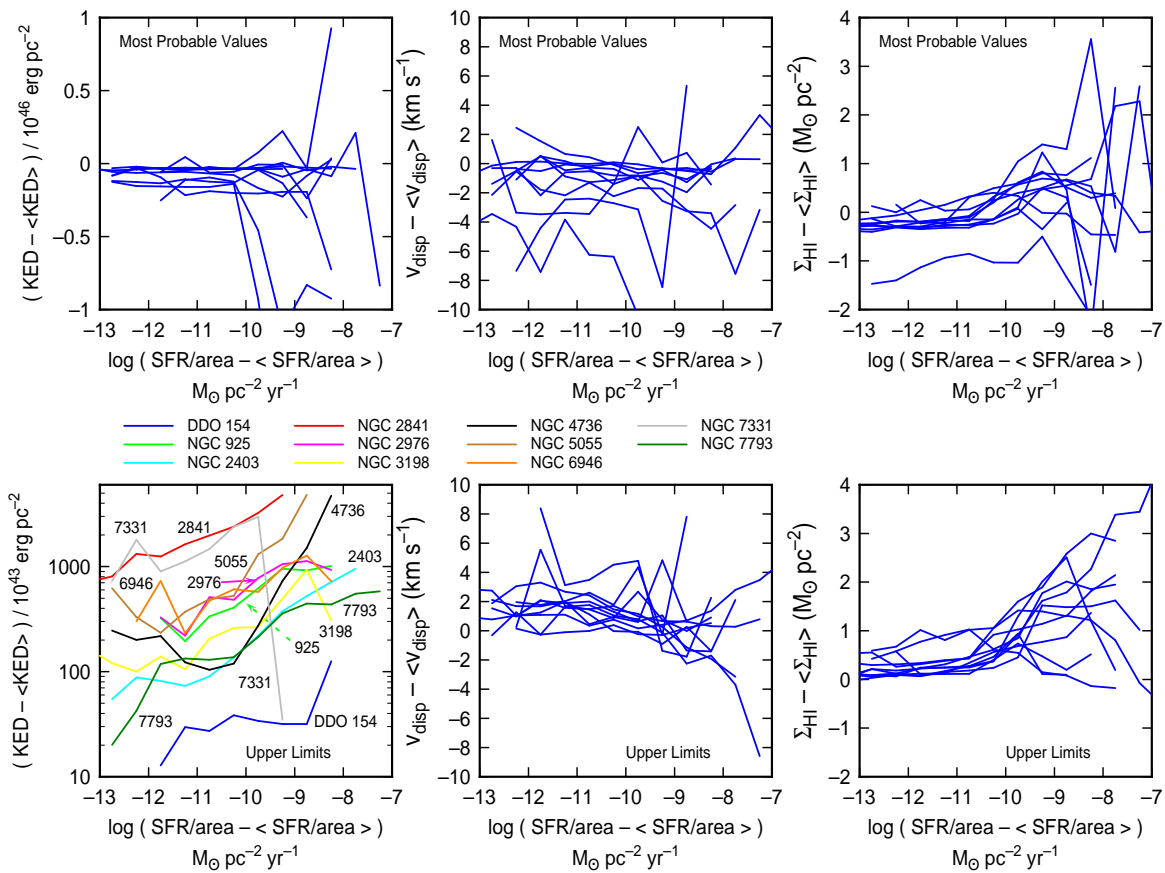}
\vskip -.8truein
\caption{(upper panels) Trend lines for the values of the three quantities at the
peaks of the pixel distributions in Figure \ref{pixel_plots_KED_vdisp_NHI}
(solid lines in Fig. \ref{KED_vdisp_NHI_peaks}),
plotted for all galaxies. (lower panels)
Statistical upper limits from the dotted curves in Figure \ref{KED_vdisp_NHI_peaks}
for all galaxies.
The upper limits for the excess KEDs are plotted on a log scale with galaxy names
indicated and also distinguished by color (only the positive values of excess KED are
included because of the log).
The units for excess KED are $10^{46}$ erg pc$^{-2}$ in the top panel and $10^{43}$ erg pc$^{-2}$
in the bottom panel, $V_{\rm disp}$ is km s$^{-1}$, $\Sigma_{\rm HI}$ is
in M$_{\odot}$ pc$^{-2}$ and excess SFRD is in M$_{\odot}$
pc$^{-2}$ yr$^{-1}$.
The distributions indicate that the excess KED (top left)
is slightly negative for most SFRDs, while the statistical
upper limit increases slightly
with SFRD (lower left). The excess velocity dispersion is also
slightly negative (upper middle) and the upper limit decreases with increasing SFRD (lower middle).
The
excess \HI\ is sightly negative for small excess SFRD with a slightly
increasing trend, and the statistical upper limit to the excess \HI\ is
positive and increases
with SFRD. One implication of these trends is that the local \HI\ surface density
decreases
slightly or stays approximately the same in a region of star formation compared to the
average value at that radius. Also, the velocity dispersion decreases in
regions of star formation, suggesting a slight cooling trend during a
conversion to molecules. There is no evidence that feedback
from star formation generates turbulence in the local HI gas.
\label{ked_vdisp_nhi_vs_sfra_6panel_with_log}}
\end{figure}

\newpage
\begin{figure}[t!]
\epsscale{1.}
\vskip -3.truein
\plotone{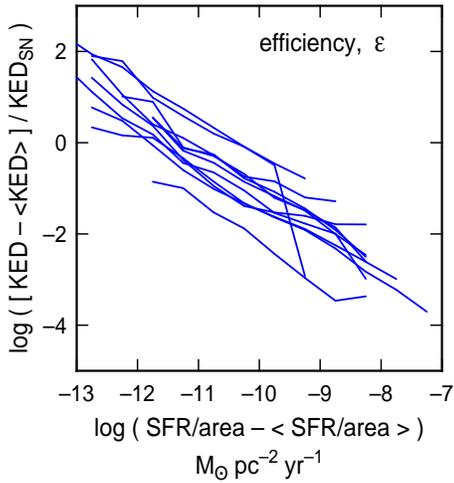}
\vskip -.7truein
\caption{The dimensionless ratio of the upper limit to the excess KED
(from the lower left panel of fig. 10) to the hypothetical KED
that would come from the excess SFRD at 100\%
efficiency for supernovae, is plotted versus the excess SFRD.  The inverse trend
arises because the upper limit to the excess KED varies more slowly than the excess SFRD.
\label{all_ked_over_KED_from_sfra_vs_sfra}}
\end{figure}

\clearpage
\appendix \label{append:A}
\section{Extended Figures}

\begin{figure}[ht]
\includegraphics[width=16cm]{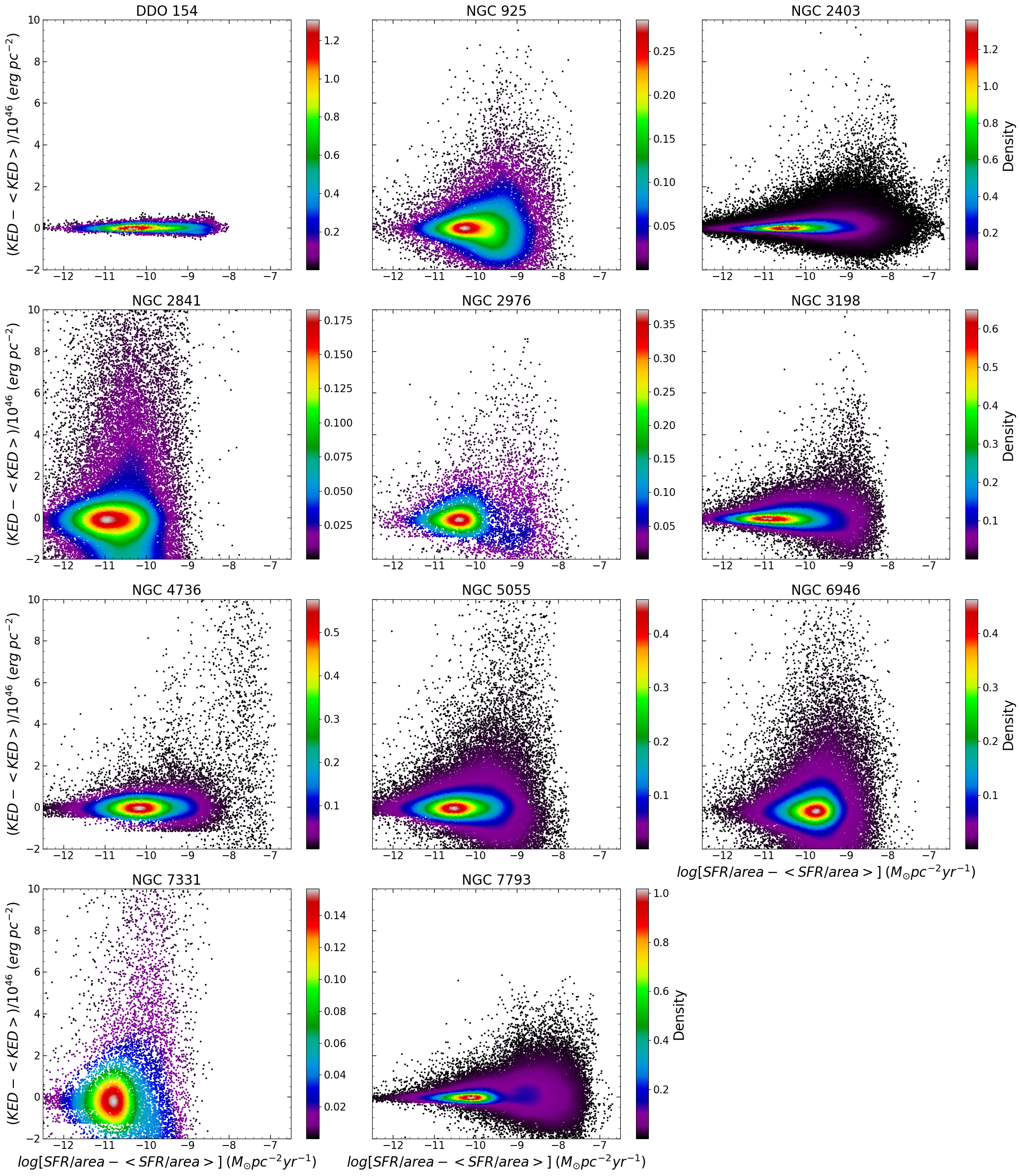}
\caption{Pixel-pixel plots showing the relationship between the excess KED in
units of $10^{46}$  erg pc$^{-2}$ and the log of the
excess SFRD in units of M$_{\odot}$
pc$^{-2}$ yr$^{-1}$. The color scale represents
the density of points. 
\label{allkedpix_v2}}
\centering
\end{figure}

\begin{figure}[ht]
\includegraphics[width=16cm]{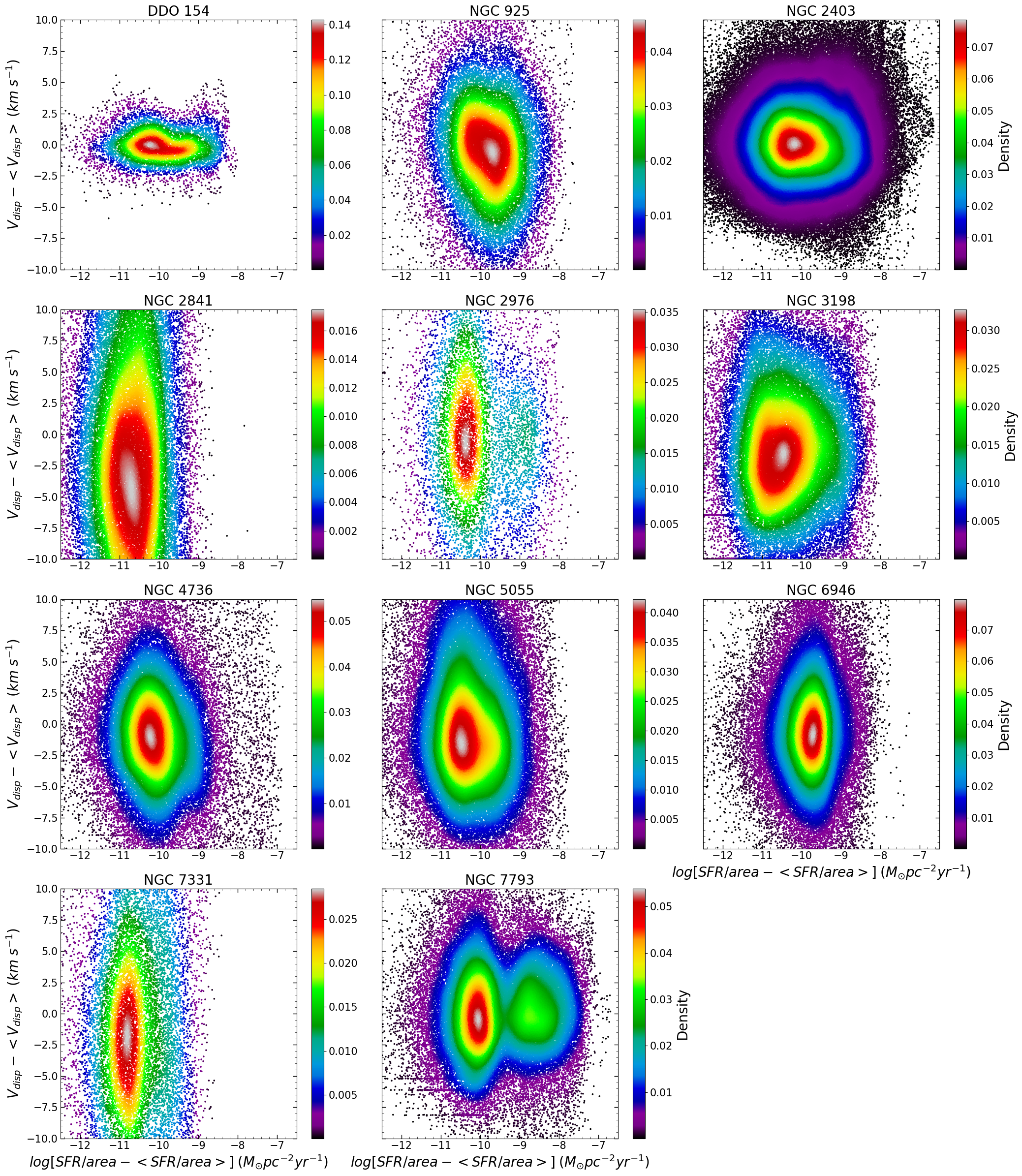}
\caption{Pixel-pixel plots showing the relationship between the excess $V_{\rm disp}$
in units of km s$^{-1}$
and the log of the excess SFRD in units of M$_{\odot}$
pc$^{-2}$ yr$^{-1}$. The
color scale represents the density of points. 
\label{allvdisppix_v2}}
\centering
\end{figure}

\begin{figure}[ht]
\includegraphics[width=16cm]{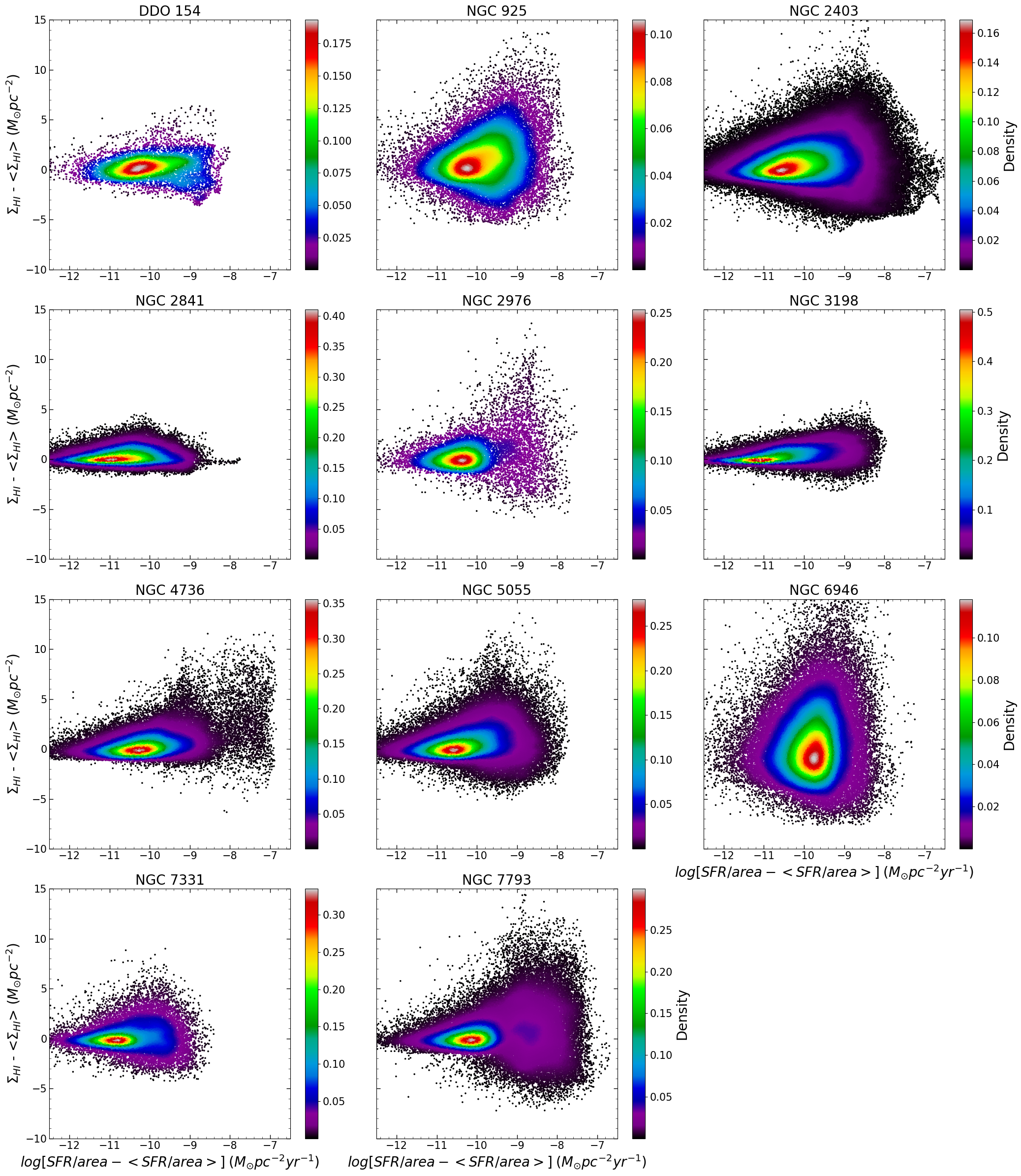}
\caption{Pixel-pixel plots showing the relationship between the excess $\Sigma_{\rm HI}$
in units of $M_\odot$ pc$^{-2}$
and the log of the excess SFRD in units of M$_{\odot}$
pc$^{-2}$ yr$^{-1}$. The color scale represents the density of points. 
\label{allsigmahipix_v4}}
\centering
\end{figure}

\end{document}